\documentclass[12pt]{book}
\usepackage[utf8]{inputenc}
\usepackage[english]{babel}
\usepackage{amssymb,amsmath}
\usepackage{caption}
\usepackage{helvet}
\usepackage{geometry}
\usepackage{graphicx}
\usepackage{authblk}
\usepackage{hyperref}
\usepackage{fancyhdr}
\usepackage{titlesec}
\usepackage{lipsum} 
\usepackage[sectionbib]{chapterbib}

\renewcommand{\thechapter}{}
\titleformat{\chapter}[hang] {\normalfont\huge\bfseries\filcenter} 
{\thechapter}{0em} 
  {}
  
\begin{document}

\chapter{Windsurf-mimetic study about unsteady propulsion}

\begin{center}
\noindent
\textbf{G. Bertrand}, \textbf{B. Thiria}, \textbf{R. Godoy-Diana}, \textbf{M. Fermigier}\\[1em]
\textit{
Laboratoire de Physique et M\'ecanique des Milieux H\'et\'erog\`enes (PMMH), CNRS UMR 7636, ESPCI Paris---Universit\'e PSL, Sorbonne Universit\'e, Universit\'e Paris--Cit\'e, 75005 Paris, France}
\end{center}

{\large \bf Abstract}\\
We study experimentally a a three-dimensional reduced model of a sail shape performing pitching oscillations around a mean incidence angle ($\alpha_{m}$) with respect to an incoming flow in a hydrodynamic channel at a constant velocity where the Reynolds number based on the mean chord of the sail is Re$_{c} = \rho U_{\infty} c / \mu = 11900$. The problem is inspired by the "pumping" maneuver used by windsurf athletes. At the start of a race or in light winds, to get or keep the board in foiling mode, for example after a tack change, athletes use intermittent propulsion by "pumping" the sail, i.e. periodically changing the angle of incidence of the sail relative to the wind. The flapping or pitching parameters and position of the sail according to the flow (incidence angle) influence the aerodynamic forces acting on the sail by destabilising the flow and generating unsteady forces. We experimentally characterise the aerodynamic forces of the sail. We compare the sailing ($C_{drive}, \ C_{drift}$) and aerodynamic ($C_{drag}, \ C_{lift}$) coefficients between a static and an oscillating sail for different flapping parameters and different mean incidence angles of the sail and angles of attack of the boat. Thanks to the use of "pumping", we observe that it is possible to generate a drive force greater than the one generated without oscillation. Furthermore, "pumping" increases the range of mean incidence angle in which the drive force is positive. However, this increase inevitably comes with an increase in drift force, which is often detrimental. These data can be used to improve the Velocity Prediction Programme (VPP) associated with windsurfing and to help athletes optimise their "pumping".

\section{Introduction}
The competitive practice of sailing and windsurfing has seen a recent revolution with the introduction of new appendages of hydrofoils that generate lift and keep the board or the boat out of the water at sufficiently high sailing speeds ($U_{\mathrm{boat}} \sim 3$ m/s) \cite{mok2023performance}. This allows one to increase the speed significantly because the wave drag and the hydrodynamic drag are almost suppressed. These innovations include the new iQFOil class introduced for the 2024 Olympic Games.

\begin{figure}[t!]
    \centering
    \includegraphics[width=0.9\linewidth]{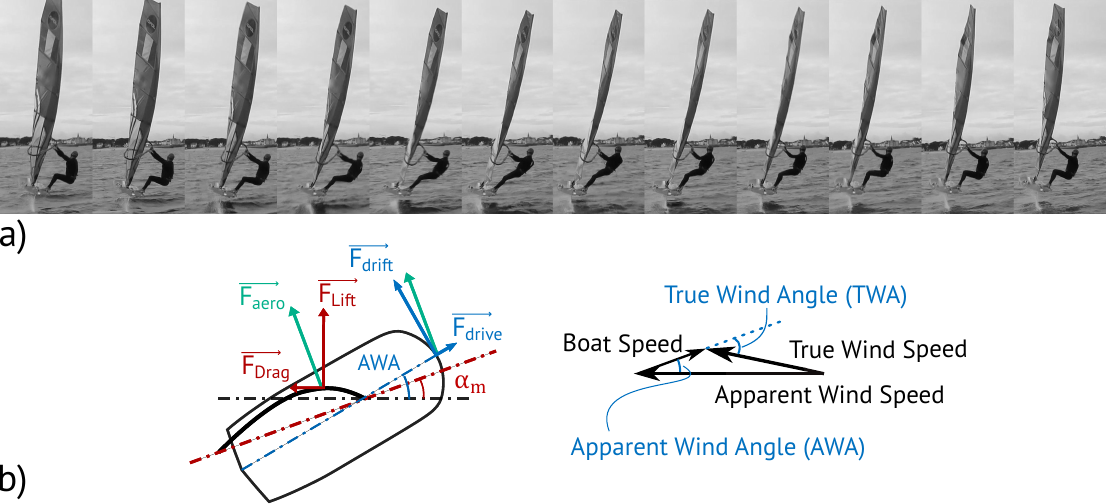}
    \caption{a) Chronophotographic sequence of a one-period "pumping" maneuver performed by a French athlete during a training session in Quiberon, 2021. From left to right, the decomposition of the athlete's movement shows how their center of mass motion enables sail oscillations. The athlete's arms are not the main cause of the movement. b) Sketch of sailboat dynamics in upwind conditions with the associated sailing speed triangle, showing forces applied to the sail and foil. The true wind angle (TWA) is defined as the angle between the true wind and the boat's longitudinal axis (blue dotted-dashed line). The aerodynamic force (green) is decomposed in both the boat reference frame (blue) and the flow reference frame (red). The apparent wind angle (AWA) is the angle between the apparent wind speed (AWS, black dashed line) and the boat speed (BS, blue dashed line) directions. $\alpha_{m}$ is the mean angle of attack: the angle between the sail's centerline (red dashed line) and the apparent wind speed direction (AWS, black dashed line). (Figure adapted from~\cite{bertrand2025}).}
    \label{pumpseq}
\end{figure}

During race starts or in low wind conditions, particularly after maneuvers like tacking which consist of turning the bow toward and through the wind to go upwind (Figure~\ref{pumpseq}.a, .b), windsurfing athletes employ a technique called "pumping" to initiate or maintain foiling where the hydrofoil lifts the board above the surface (Figure~\ref{pumpseq}.a). This involves rhythmically adjusting the sail's angle relative to the wind through an up and down movement of the athlete's centre of mass, causing the sail to oscillate and provide intermittent propulsion to keep the board above water (Figure~\ref{pumpseq}.a).

Few studies have been conducted to examine the performance of a "pumping" sail according to complex parameters for windsurfing and sailing using a symmetrical foil \cite{zhou2021propulsive, Aubin:2016, Young:2019, bertrand2025}. These studies analysed the efficiency and the unsteady drive force, which is the aerodynamic force projected in the boat traveling direction (Figure~\ref{pumpseq}.b), according to flapping parameters and sailing kinematics parameters. These studies show that unsteady propulsion methods are effective in upwind conditions (Figure~\ref{pumpseq}.b). When windsurfing athletes perform the "pumping" motion, the resulting sail kinematics is three dimensional (3D), but a reasonable leading-order model  can be obtained by limiting the motion to a rotation around the vertical axis. In the present work, we will use such a simplified model: a pitching rigid sail. In windsurfing, athletes navigate under various wind conditions, at incidence angles ranging from small to large where Leading-Edge Vortices (LEVs) may form, increasing the lift force on the sail and delaying the dynamic stall \cite{ohmi1990vortex, Ohmi:1991, schouveiler2005performance, Eldredge:2010}.

Using videos from training sessions of the French sailing team we were able to calculate the order of magnitude of the Strouhal number during "pumping" maneuvers, a non-dimensional number defined as the ratio of the beating speed of the sail over the flow velocity (St$_{A} = fA/U_{\infty}$, $f$ is the pitching frequency and $A$ the beating amplitude). The athletes pump the sail with a frequency $f \sim 1$ Hz and a beating amplitude approximatively equal to the half of the board width $A \sim 0.5$ m. The flow velocity perceived by the sail in the lowest use range is $U \sim 5$ m/s. These physical parameters of the "pumping" kinematics give St$_A \sim 0.1$. In this study, St$_A$ will be varied in the range between 0 and 0.22 (Table~\ref{tabvoile}). The orders of magnitude of wind speed, ranging from 5 m/s to 15 m/s, enable the calculation of the Reynolds number associated with the sail chord $c$, Re$_{c} = U_{\infty} c / \nu$, where $U_{\infty}$ is the flow velocity perceived by the sail and $\nu$ is the kinematic viscosity of the fluid. In windsurfing, the range of the Reynolds number is between $\mathrm{10^5}$ and $10^6$ \cite{mok2023performance}.

We define the true wind angle (TWA) as in Figure~\ref{pumpseq}.b to characterise how the boat is moving according to the true wind direction, consistent with nautical studies. We focused on upwind sailing condition (TWA~$<$~90$^{\circ}$). A sketch of the balance of forces applied on the sail according to the direction of the wind is shown in Figure~\ref{pumpseq}.b. The boat has a driving direction given by the boat speed (BS). The apparent wind speed (AWS) is the composition of the BS and the true wind speed (TWS) and represents the wind perceived by the boat. The apparent wind angle (AWA) is the angle between BS and AWS. The sail produces aerodynamic forces depending on the incidence angle ($\alpha_{m}$), which is the angle between the sail's centerline (red dashed line) and the apparent wind speed direction (AWS, black dashed line). AWA and $\alpha_{m}$ are independent, but both of them influence the drive and the drift forces and so the performance of the boat (Figure~\ref{pumpseq}.b).

In this study, the impact of "pumping" amplitude and frequency, as well as the influence of mean incidence angle, will be examined by comparing measurements of lift and drag forces on a reduced-scale rigid sail with a complex shape. This investigation is done according to a real windsurf sail shape, where the intrados and extrados are different, in order to mimic the behaviour of a real sail under real sailing conditions, especially upwind conditions (Figure~\ref{pumpseq}). Subsequently, the forces within the boat's frame of reference (drive and drift forces) are studied in order to characterise the sailing performance of a windsurf sail in steady and unsteady modes. 

\section{Methodology}
We aim at characterising the performance of the unsteady propulsion of a windsurf sail experimentally. Using the 3D scan of an iQFoil sail in static conditions (without wind flow) at the Ecole Nationale de Voile et des Sports Nautiques (ENVSN, Quiberon), and the digital modifications made to the points cloud by Benoît Augier and his team at IFREMER (Brest) to approximate a dynamic sail shape, we obtained a realistic sail model where a twist angle of $20$ degrees between the top and bottom of the sail was artificially added. We present in Figure~\ref{fig:drafting_sail} different views of this sail.

\begin{figure}[!h]
    \centering
        \includegraphics[width=0.9\linewidth]{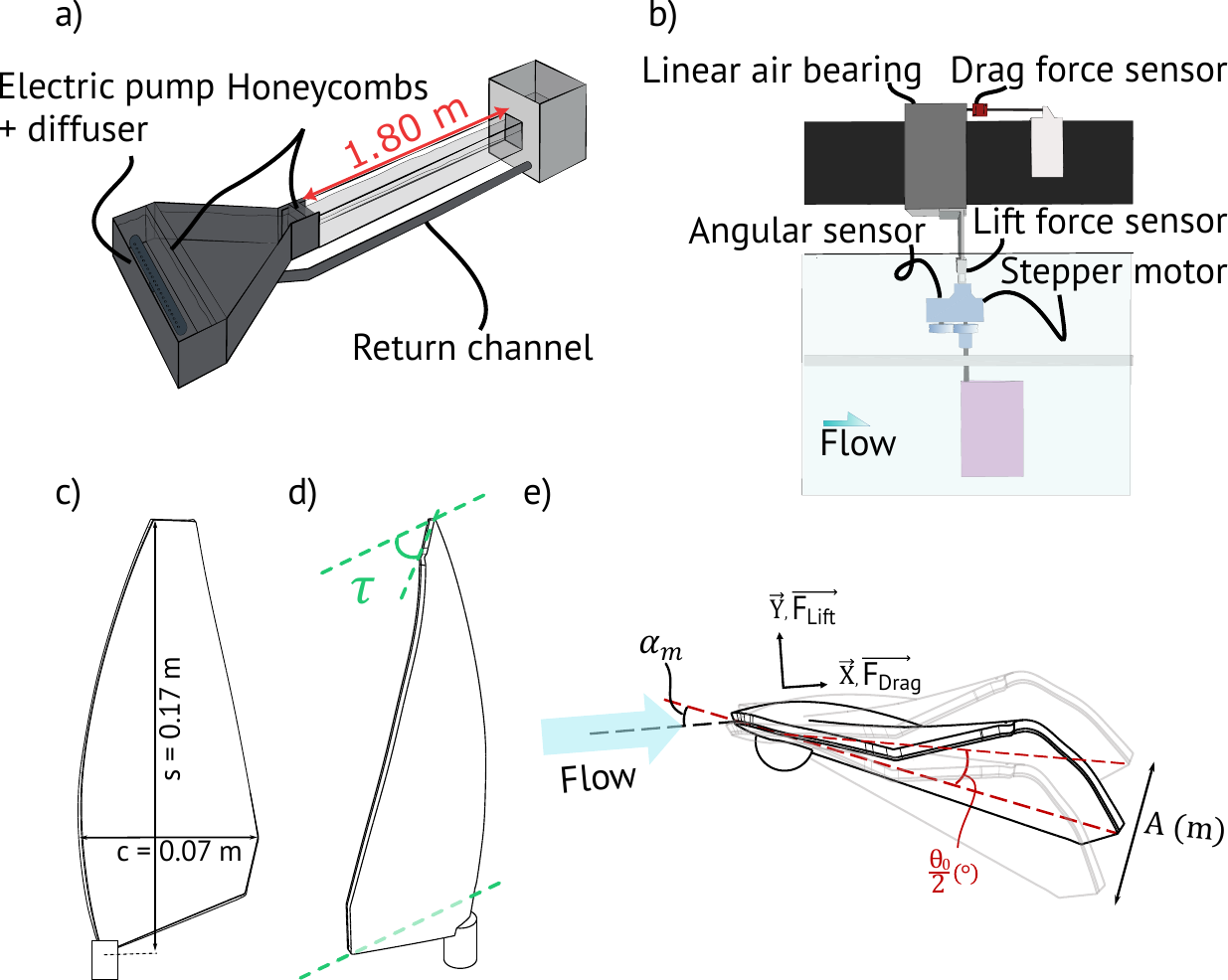}
    \caption{Sketches of the setup and technical drawings of the sail on a scale of 1/30, where we added a perforated cylinder to the base of the mast, which allows us to attach the sail to our force measurement system. a) Closed loop water channel with a length of $1.80$ m and a cross-section with water of $0.2 \times 0.2$ m$^{2}$ \cite{auregan2023scaling}. b) Control and acquisition setup to measure forces and kinematics data. c) Side view for defining the chord $c$ and the span $s$ of the sail. d) Isometric rear view for defining the twist angle $\tau$ between the top and bottom of the sail, which is $20^{\circ}$ in this case. e) Sketch in top-view of an experiment of pitching sail.}
    \label{fig:drafting_sail}
\end{figure} 

A ratio of 1/30 was applied so that it could be tested in a hydrodynamic channel. The chord c is defined by the horizontal segment passing through the clew such that c = 0.07 m and a span s = 0.17 m, which gives an aspect ratio AR = 2.43 (Figure~\ref{fig:drafting_sail}.c). The scaled sail was 3D printed in PLA, making it rigid.
Experiments were performed in a free surface and closed loop water channel with a $0.2$ m by $0.2$ m section. The turbulence intensity measured by Particle Image Velocimetry is below $5\%$ \cite{auregan2023scaling}.

In order to study the effect of unsteady propulsion, we conducted pitching experiments (Figure~\ref{fig:drafting_sail}.e) over a wide range of mean incidence angles (or angles of attack) $\alpha_{m}$ up to 35$^{\circ}$ by changing it in increments of 5$^{\circ}$. We chose the range of St$_{A}$ to test, here going up to $0.22$ in increments of $0.06$, as well as the frequency range to study, going from $1$ Hz to $3$ Hz in increments of $0.5$ Hz. We define here the reduced frequency $k = 2 \pi f c / U_{\infty}$, as a non dimensional frequency (Table~\ref{tabvoile}) For each frequency value stated, we explore the complete range of St$_{A}$. We therefore carry out experiments with 4 different amplitudes per frequency. The following experiments were performed at a flow velocity U$_{\infty} = 0.17$ m/s corresponding to a Reynolds number based on the chord Re$_c$ = 11900. We summarise the experimental data in Table~\ref{tabvoile}.

The "pumping" maneuver is reduced to a pitching motion characterized by the angle between the flow direction and the foil chord as $\theta(t) = \alpha_{m} + (\theta_{0}/2)\sin{(2 \pi f t)}$, where $\alpha_{m}$ is the mean incidence angle, $f$ the frequency of pitching and $\theta_{0}$ the beating angular amplitude. The amplitude swept by the trailing edge is $A = (2c)\sin{(\theta_{0}/2)}$ (Table~\ref{tabvoile}).

Sketches of the water channel and of the experimental setup are presented in Figure~\ref{fig:drafting_sail}.a, .b. The rotation axis for the pitching movement is located at $0.1$c from the leading edge. The axis is a carbon rod attached to a stepper motor that also drives an angular position sensor. The ensemble rotation motor + foil is mounted on a load sensor (CLZ639HD, measuring up to $\pm 0.98$ N) that measures the lift component (Y) of the force. This system previously presented is in a sliding connection with an air cushion in relation to the frame. A second load sensor (FUTEK LSB210 measuring up to $\pm 1$ N) working in traction and compression, located between the linear air bearing and the frame, records the drag force. We sample analogically all the physical parameters and we control the command sent to the stepper motor with a National Instruments card (NI-USB-6221) which allows us to record data at a frequency of 1024 Hz after passing through an amplifier.

By controlling the frequency, amplitude and mean incidence angle of the sinusoidal motion of the sail, and thanks to the two force sensors (Figure~\ref{fig:drafting_sail}.c), we are able to study unsteady propulsion as a function of the physical and kinematic parameters of the pitching motion described in Table~\ref{tabvoile}.

\begin{center}
\begin{tabular}{|c||c||c||c||c|}
\hline Re$_{c}$ & $\alpha_{m}$($^\circ$) &  $k$ & St$_{A}$ &  A (m)\\
11900 &  [\![-5 , 35]\!] &  [1.3 , 3.9] & [0.06 , 0.22]  &  [0.0015 , 0.018] \\
\hline 
\end{tabular}
\captionof{table}{Physical parameters describing our experiments. The study is performed for a constant flow velocity $U_\infty = 0.17$ m/s.}
\label{tabvoile}
\end{center}

We run experiments for each case during 30 cycles. We present in Figure~\ref{fig:drafting_sail} an example of raw force measurements and filtered force measurements.

\begin{figure}[ht]
    \centering
        \includegraphics[width=1\linewidth]{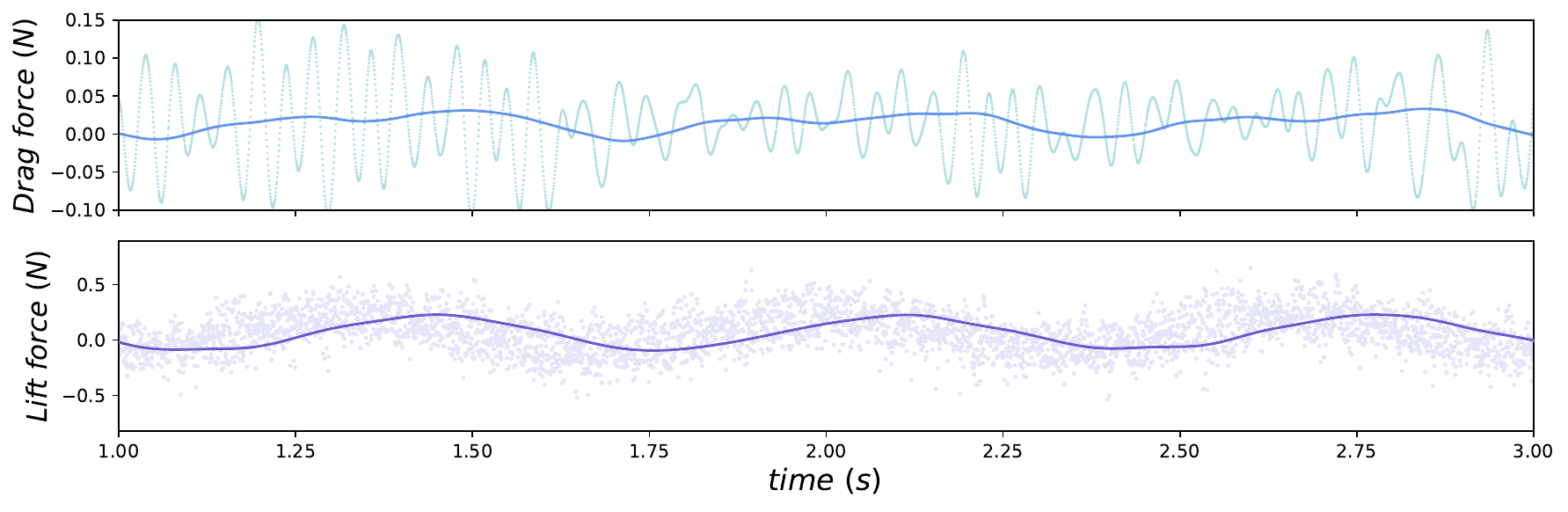}
    \caption{Raw (points) and filtered (lines) data of drag (top) and lift (bottom) forces versus time, for an experiment where $\alpha_{m}$ = 5$^\circ$ , $\theta_{0}$ = 14$^\circ$ and $f = 1.5$ Hz ($St_{A}=0.18$). We have smoothed here with a low-pass filter (Butterworth filter). We can observe the phase between the raw signal and the filtered signal for each case. The lift force is periodic with $f_{Lift} = f$. The drag force signal is periodic too with $f_{Drag} = 2f$.}
    \label{fig:drafting_sail}
\end{figure}

We measure the mean value of lift and drag for the couple (St$_{A}$, $\alpha_{m}$). We run experiments for each case during 30 cycles, and we extract the mean value of forces. We define the aerodynamic coefficients classicaly for high Reynolds number flow as, $C_{D}=\overline{F_{Drag}} / \frac{1}{2} \rho S U_{\infty}^{2} \ \ \text{and} \ \ C_{L}=\overline{F_{Lift}} / \frac{1}{2} \rho S U_{\infty}^{2}$, where $C_{L}$ and $C_{D}$ represent the lift and the drag coefficient respectively, where $\rho$ is the fluid density (1000 kg/m$^{3}$), $S$ is the lifting surface (0.01 m$^{2}$), $U_{\infty}$ is the flow velocity (0.17 m/s) and $F_{Drag}$ and $F_{Lift}$ the drag and the lift force, respectively. Due to the size of the water tank, a correction for the blockage effect is taken into account. For our experimental set-up the blockage area ratio $S_{a}/S_{tank}$ is up to $10\%$, where $S_{tank}$ is the cross flow surface of the water tank ($S_{tank} = 0.04$ m$^{2}$) and $S_a$ is the projected surface orthogonal to the flow of the foil depending on the mean incidence angle. We use a correction with a quasi-streamlined flow method for three-dimensional bluff-body changing its mean incidence angle according to the ESDU Technical Committees \cite{esdu:76028} and based on the work of Souppez et al. \cite{souppez2022high} on a thin arc-shaped profile,
\begin{align}
    \frac{C_{*,c}}{C_{*}} = 1 - \lambda_{1}\lambda_{3}\lambda_{5} \left( 1 + 0.51 \frac{\sqrt{S_a}}{c} \right) \frac{c S_a}{S_{tank}^{1.5}} - 0.5\frac{S_a}{S_{tank}} C_{D},
\end{align}
$C_{*,c}$ represents the force coefficient corrected and $C_{*}$ the raw coefficient. $\lambda_{1} = 0.72 \times (l/h + h/l)$ is the water tank shape parameter for a three-dimensional flow. $l$ and $h$ are respectively the width and the height of the water tank. $\lambda_{3}= \frac{V_{sail}}{cS_{a}}$ is the body volume parameters. $\lambda_{5} = 1 + 1.1 \times (c/w) \times(\pi/180)^{2} \times \alpha_{m}^{2}$, for taking into account the projected width when $\alpha_{m}$ in degree changes. $w$ is the maximum width of the foil.

\section{Results and Discussion}

\begin{figure}[!t]
    \centering
        \includegraphics[width=1\linewidth]{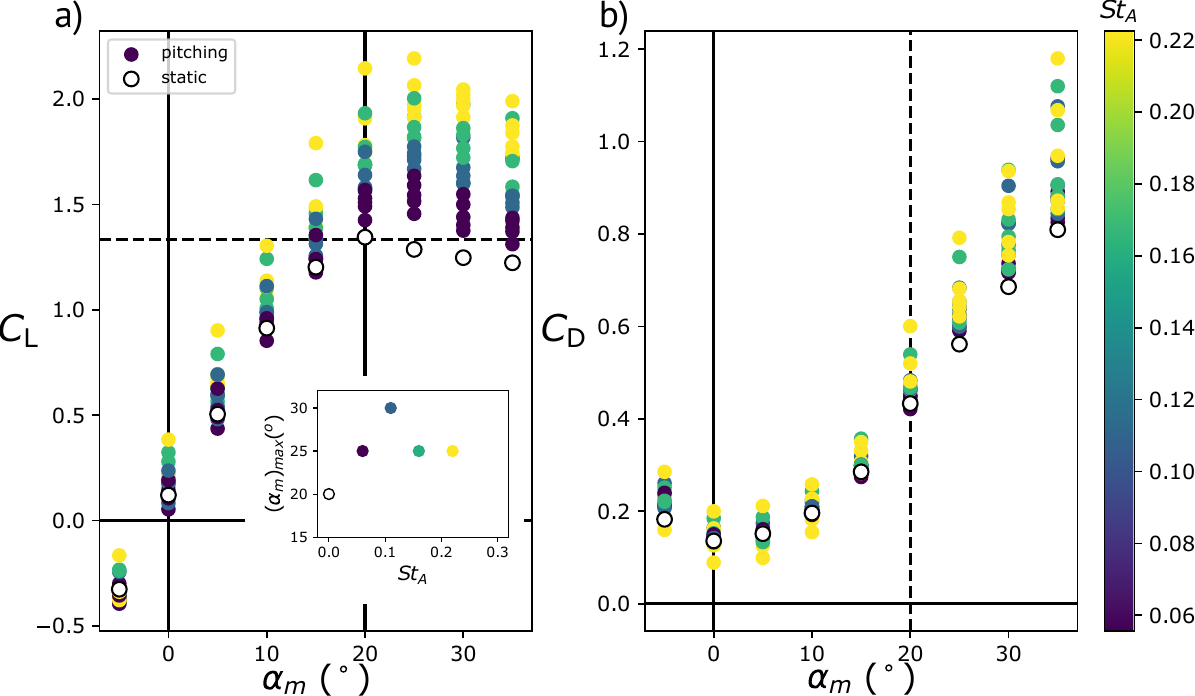}
    \caption{Lift and drag coefficients as a function of $\alpha_{m}$ for a range of Strouhal numbers (colorbar) such as $St_{A} = [0, 0.06, 0.12, 0.18, 0.22]$ (Table~\ref{tabvoile}). a) Lift coefficients. A vertical line crosses $\alpha_{m} =$ 20$^\circ$ where the static stall appears. In insert, we show the value of $\alpha_{m}$ where the maximum of lift coefficient is reached as a function of St$_{A}$. b) Drag coefficients.}
    \label{fig:pitch_polar_sail}
\end{figure}

Figure~\ref{fig:pitch_polar_sail} compares the experimental results of the unsteady forces (coloured dots) measured on the rigid sail with the static measurements (white hollow dots) in the form of a polar plot of the coefficients $C_{L}$ and  $C_{D}$ as a function of the mean incidence angle $\alpha_ {m}$.

$\bullet$ $C_{L}$, $0^{\circ} < \alpha_{m} < 20^{\circ}$:
For $\alpha_{m} = 0^{\circ}$, the static coefficients $C_{L}$ and $C_{D}$ are greater than 0. This is due to the asymmetrical geometry of the wing. The static lift polar curve shows that static stall occurs at $\alpha_ {m} \approx 20^{\circ}$ for a value of $C_{L} = 1.34$. Comparing to the measurements for a symmetrical NACA0018 profile reported in \cite{bertrand2025}, this corresponds to an increase of approximately $0.44$ in the lift coefficient. The oscillation of the sail will generate a greater lift force than in static conditions in almost all cases of St$_{A}$ studied. However, in this range of $\alpha_{m}$ and for a fixed mean incidence angle, St$_{A}$ is not sufficient to characterise the behaviour of $C_{L}$ and fit the experimental data. Oscillating the sail generates a variation in $C_{L}$ over a range of approximately 0.5 in coefficient value.

\begin{figure}[!t]
    \centering
        \includegraphics[width=1\linewidth]{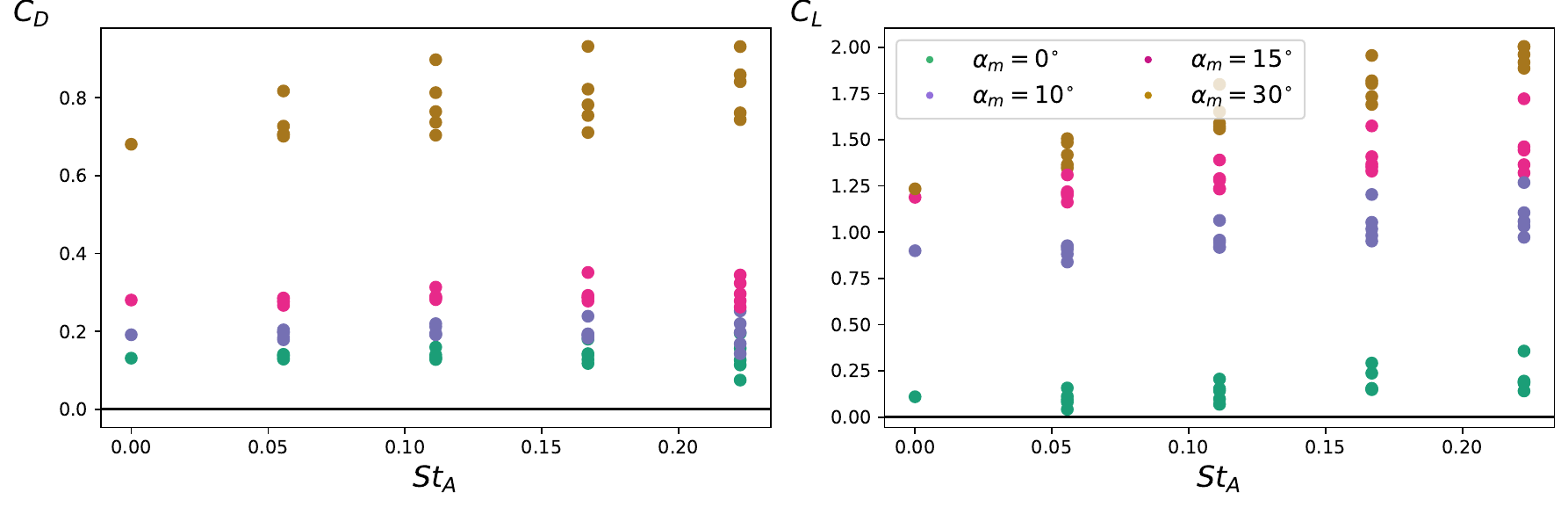}
    \caption{Aerodynamic force coefficients versus the Strouhal number for 4 values of mean incidence angle. Left: $C_{D}$ vs St$_A$. Right: $C_{L}$ vs St$_A$. For each value of St$_A$, we performed experiments with different pairs ($f$, $A$).}
    \label{fig:clcd_vs_Sta}
\end{figure}

$\bullet$ $C_{L}$, $\alpha_{m} \geq 20^{\circ}$:
At average incidence angles greater than the static stall angle $\alpha_{m, static \ stall} \approx 20^{\circ}$, oscillating the sail generates more lift and also delays the dynamic stall. Increasing St$_{A}$ does not seem to delay the stall any further, as shown by the values of $(\alpha_{m})_{max}$ in the inset  of Figure~\ref{fig:pitch_polar_sail}a. In this interval of $\alpha_{m}$, St$_{A}$ seems to be an appropriate parameter to describe the trend of $C_{L}$: at each value of $\alpha_m$, $C_{L}$ increases monotonically with increasing St$_A$ over a range going up to values $\approx 60\%$ higher than the static stall value (Figure~\ref{fig:clcd_vs_Sta}).

$\bullet$ $C_{D}$:
In static conditions, $C_{D}$ typically increases proportionally to $\alpha_{m}^{2}$. No clear trend in the variation of drag force coefficients is observed with respect to St$_A$ (Figure~\ref{fig:clcd_vs_Sta}). The influence of pitching, in our range of measurements, will impact $C_{D}$ in a range of approximately +0.15 in coefficient compared to the static value, for $0^{\circ} \leq \alpha_{m} < 20^{\circ} $, then by approximately 0.25 for $20^{\circ} \leq \alpha_{m} < 35^{\circ}$ and finally reaching approximately 0.4 for $\alpha_{m} = 35^{\circ}$. In a few cases for $\alpha_{m} < 10^{\circ}$ a decrease in the drag coefficient with respect to the static value is measured.

These behaviours directly influence the boat's drive and drift forces. Using Figure~\ref{pumpseq}.a, we define the sailing force coefficients $C_{drive}$ and $C_{drift}$ as,
\begin{align}
    C_{drive}= C_{L}(\alpha_{m}) \sin{(\mathrm{AWA})} - C_{D}(\alpha_{m}) \cos{(\mathrm{AWA})}, \label{drive_coeff}
\end{align}
\begin{align}
    C_{drift}= C_{L}(\alpha_{m}) \cos{(\mathrm{AWA})} + C_{D}(\alpha_{m}) \sin{(\mathrm{AWA})}, \label{drift_coeff}
\end{align}

\begin{figure}[!t]
    \centering
        \includegraphics[width=1\linewidth]{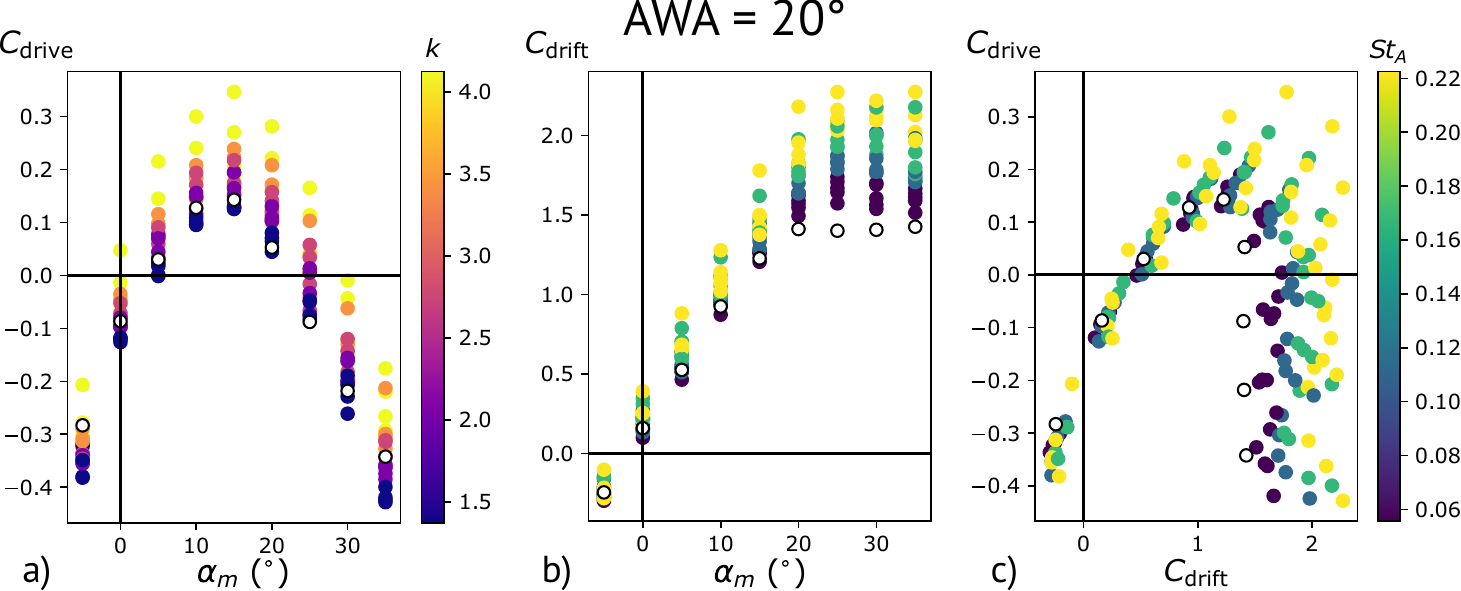}
    \caption{Sailing force coefficients as a function of $\alpha_{m}$ and sailing polar of drive and drift coefficients for AWA $=20^{\circ}$. a) $C_{drive}$ vs $\alpha_{m}$ scaled with $k$, b) $C_{drift}$ vs $\alpha_{m}$ scaled with St$_A$, c) $C_{drive}$ vs $C_{drift}$ scaled with St$_A$.}
    \label{fig:polars_various_awa}
\end{figure}

Figure~\ref{fig:polars_various_awa} shows the polars of the forces projected in the boat's reference frame $C_{drive}$ and $C_{drift}$ for AWA = 20$^{\circ}$ representing a classical mean value for upwind sailing in iQFoil. 

We plot $C_{drive}$ as a function of $\alpha_m$ sorted by the reduced frequency $k$, which gives a correct trend for the evolution of the coefficient for a fixed mean incidence angle. According to Figure~\ref{fig:polars_various_awa}.a oscillating the sail will allow a transition from a non-propulsive situation to a propulsive situation, such as when AWA = 20$^{\circ}$, from $\alpha_m$ = 0$^{\circ}$ to $\alpha_m$ = 25$^{\circ}$. The maximum $C_{drive}$ reaches approximately 0.35 for $\alpha_m$ = 15$^{\circ}$. For a given value of $\alpha_m$, the dispersion of $C_{drive}$ is almost consistent with $k$. Figure~\ref{fig:polars_various_awa}.a clearly shows that the range of propulsion is increased by "pumping". In relation to Figure~\ref{fig:polars_various_awa}.b, $C_{drift}$ generated by pitching or in static conditions increases linearly up to static stall and then appears to become constant. For $\alpha_m <$ 20$^{\circ}$, there is no consistency between the increase in $C_{drift}$ for an increase in St$_A$. However, for $\alpha_m \geq$ 20$^{\circ}$, the increase in $C_{drift}$ is related to the increase in St$_A$.

Finally, the force polar is shown in Figure~\ref{fig:polars_various_awa}.c. This data can be supplemented with measurements taken on the hydrofoil of the windfoil to approximate the overall effect of "pumping". Oscillating the sail inevitably generates a non-stationary drifting force. For this classical case, when AWA = 20$^{\circ}$, by "pumping", athletes can increase the range of $\alpha_m$ where the drive force is positive, between 0$^{\circ}$ and 25$^{\circ}$. In situations such as starting or tacking, we can assume that during the short time interval when athletes oscillate the sail, the main objective is to increase the drive force in order to increase the speed of the boat and make it fly above the water.

\section{Conclusion}

We have presented the operating ranges of sail force coefficients for a profile with the geometric characteristics of a sail (3D shape, heterogeneous curvature, twisted angle). The study of the pitching sail using wide ranges of kinematic parameters show us, in a first order, a complex dynamics. The pursuit of this study will allow us to identify strategies for implementing "pumping" in accordance with the prevailing sailing conditions and, maybe, help athletes.

\section{Acknowledgement}
This work was supported by the French National Research Agency with grant "Sport de Très Haute Performance" ANR 19-STHP-0002, "Du Carbone à l'Or Olympique" (CtoOr) project. The authors would like to thank Amaury Fourgeau for technical support. We are grateful to Benoît Augier for the three-dimensional shape of the sail and also for fruitful discussions and scientific suggestions.


\begingroup
\makeatletter
\renewcommand\chapter{\@startsection{section}{1}{\z@}{-3.5ex plus -1ex minus -.2ex}{2.3ex plus .2ex}{\normalfont\Large\bfseries}}
\bibliographystyle{unsrt}
\bibliography{biblio.bib}
\makeatother
\endgroup

\end{document}